# Multiferroics seen from theoretic derivation of a tight binding model

T. D. Cao

*Department of physics, Nanjing University of Information Science & Technology, Nanjing 210044, China*

One presented some lattice models, while the theoretic derivation has not been found, and the importance of correlation effects has to be emphasized. On the basis of the non-relativistic Hamiltonian from the Dirac equation, we derive in detail a tight binding model. We find that both ferromagnetism and ferroelectricity are from the correlation effect between electrons, and magnetic and electric orders are strongly coupled due to the spin-orbit interaction.

Multiferroics are materials which can show both ferromagnetism and ferroelectricity, and their magnetic and electric orders are strongly coupled. Some properties of them may be included in first quantum models or phenomenological models [1,2], however, these materials have features of the strongly correlated systems, the many-body effects should be important for the changes of their magnetic and electric orders. Some many-body models have been suggested [3-6], but different materials have various features, thus a general microscopic model should be found. That is to say, some properties must be attributed to electron states and interactions within a molecule, while some properties have to be attributed to the correlation effects between electrons. Particularly, some microscopic mechanisms could be found in the theoretic derivation of a model. This work derives a tight binding model which should show these correlation effects.

To derive a tight binding model and find the meaning of each term, we first consider the non-relativistic Hamiltonian from the Dirac equation, and then we arrive at the Hamiltonian operator for electron systems in the external fields

$$\hat{H} = \hat{H}_0 + \hat{H}_{e-e} + \hat{H}_{s-L} + \hat{H}_{ext} \tag{1}$$

$$\hat{H}_0 = \sum_i [(\hat{\vec{p}}_i - e\vec{A}_i)^2 / 2m + U_i] \tag{2}$$

$$\hat{H}_{e-e} = \sum_{i,j} V(\vec{x}_i - \vec{x}_j) \tag{3}$$

$$\hat{H}_{SL} = \sum_i i\hat{\vec{\sigma}}_i \cdot [(\hat{\vec{p}}_i - e\vec{A}_i) U_i \times (\hat{\vec{p}}_i - e\vec{A}_i)] \tag{4}$$

$$\hat{H}_{ext} = \sum_i e\hbar \hat{\vec{\sigma}}_i \cdot (\vec{\nabla}_i \times \vec{A}_i)/2m = \sum_i g_s \hat{\vec{S}}_i \cdot \vec{B}_i \tag{5}$$

Where the electron-ion interaction $U_i = \sum_l U_l(\vec{x}_i) = \sum_l U(\vec{x}_i - \vec{R}_l)$ and the vector potential $\vec{A}_i = \vec{A}(\vec{x}_i)$. The interactions between electron and electric field are neglected. The operator $\hat{H}_{e-e}$ represents the electron-electron interactions, $\hat{H}_{SL}$ represents the spin-orbit interactions, and $\hat{H}_{ext}$ represents the interactions between spins and magnetic fields.

To arrive at the so-called second quantization, we should take the complete set of states $\phi_n(\vec{x}-\vec{R}_l)$ which are assumed as the states of an electron around an ion

$$[(\hat{\vec{p}}-e\vec{A})^2/2m+U(\vec{x}-\vec{R}_l)]\phi'_n(\vec{x}-\vec{R}_l)=\varepsilon_n\phi'_n(\vec{x}-\vec{R}_l) \tag{6}$$

$\vec{R}_l$ represents the position of an ion. The magnetic field may vary with positions of ions, but $\vec{A}$ is taken as a constant around a site $\vec{R}_l$, thus $\phi'_n(\vec{x}-\vec{R}_l)=e^{ie\vec{A}\cdot(\vec{x}-\vec{R}_l)/\hbar}\phi_n(\vec{x}-\vec{R}_l)$, and the "electron orbits" $\phi_n(\vec{x}-\vec{R}_l)$ meet

$$[\hat{\vec{p}}^2/2m+v(\vec{x}-\vec{R}_l)]\phi_n(\vec{x}-\vec{R}_l)=\varepsilon_n\phi_n(\vec{x}-\vec{R}_l) \tag{7}$$

For a one band model, we will neglect the index $n$ below and expand the wave function in terms of these electron orbits

$$\psi(\vec{x})=\sum_{l,\sigma}c_{l\sigma}\phi'(\vec{x}-\vec{R}_l)\chi_\sigma \tag{8}$$

where $c_{l\sigma}$ are fermions destruction operators and $\chi_\sigma$ are the spin wave functions. The part in Eq. (1) is written as $H_0=\sum_{l,l',\sigma}H_{ll'}c^+_{l\sigma}c_{l'\sigma}$ in terms of creation and destruction operators, and the derivation is shown below

$$H_{ll'}=\int d^3x\phi'(\vec{x}-\vec{R}_l)^*[(\hat{\vec{p}}-e\vec{A})^2/2m+\sum_{l''}U(\vec{x}-\vec{R}_{l''}-\delta\vec{R}_{l''})]\phi'(\vec{x}-\vec{R}_{l'})$$
$$=\varepsilon\delta_{l,l'}+\int d^3x\phi'(\vec{x}-\vec{R}_l)^*\sum_{l''(\neq l')}U(\vec{x}-\vec{R}_{l''})\phi'(\vec{x}-\vec{R}_{l'}) \tag{9}$$

$\chi_\sigma$ are assumed having been renormalized: $\int ds|\chi_\sigma|^2=1$. Because of the localization of $\phi(\vec{x}-\vec{R}_l)$, we can take the nearest-neighbor approximation and the approximation $\vec{A}_l=\vec{A}_{l+\delta}\equiv\vec{A}(\vec{R}_l)$, and obtain

$$H_0=\varepsilon\sum_{l,\sigma}c^+_{l\sigma}c_{l\sigma}+\sum_{l,\delta,\sigma}(te^{-ie\vec{A}_l\cdot\vec{\delta}/\hbar}c^+_{l\sigma}c_{l+\delta\sigma}+h.c.) \tag{10}$$

where $\vec{\delta}=\vec{R}_{l+\delta}-\vec{R}_l$.

The electron-electron interactions are written in

$$H_{e-e}=\sum_{\substack{\nu\mu\lambda\kappa\\\sigma,\sigma'}}c^+_{\nu\sigma}c_{\mu\sigma}c^+_{\lambda\sigma'}c_{\kappa\sigma'}V_{\nu\mu\lambda\kappa} \tag{11}$$

where $\phi_\kappa(\vec{x})\equiv\phi(\vec{x}-\vec{R}_\kappa)$ and $V_{\nu\mu\lambda\kappa}=\int d^3x_1\int d^3x_2\phi^*_\nu(\vec{x}_1)\phi_\mu(\vec{x}_1)V(\vec{x}_1-\vec{x}_2)\phi^*_\lambda(\vec{x}_2)\phi_\kappa(\vec{x}_2)$. The well-known part in Eq.(11) is the on-site interaction $H_{e-e}\sim U\sum_l n_{l\sigma}n_{l\bar{\sigma}}$ and the intersite interaction $\sum_{l,l',\sigma,\sigma'}V_{ll'}n_{l\sigma}n_{l'\sigma'}$, and other terms could be found.

Note $\hat{\sigma}_x\chi_\sigma=\chi_{\bar{\sigma}}$, $\hat{\sigma}_y\chi_\sigma=i\sigma\chi_{\bar{\sigma}}$, and $\hat{\sigma}_z\chi_\sigma=\sigma\chi_\sigma$ with $\sigma=\pm 1$, this leads to $\int ds\chi^*_\sigma\hat{\vec{\sigma}}\chi_{\sigma'}=\delta_{\sigma'\bar{\sigma}}\vec{e}_x+i\sigma'\delta_{\sigma'\bar{\sigma}}\vec{e}_y+\sigma\delta_{\sigma'\sigma}\vec{e}_z$. Therefore, following similar derivation of $H_0$, we obtain the spin-orbit interactions

$$H_{SL}=\sum_{l,l',\sigma}i[c^+_{l\sigma}c_{l'\bar{\sigma}}(\vec{e}_x-i\sigma\vec{e}_y)+\sigma c^+_{l\sigma}c_{l'\sigma}\vec{e}_z]\cdot\vec{H}^{SL}_{ll'} \tag{12}$$

In Eq.(12) we have introduced

$$\vec{H}_{ll'}^{SL} = i\int d^3x \phi^{*'}(\vec{x}-\vec{R}_l)[(\hat{\vec{p}}-e\vec{A})\sum_{l''}U(\vec{x}-\vec{R}_{l''}-\delta\vec{R}_{l''})]\times(\hat{\vec{p}}-e\vec{A})\phi'(\vec{x}-\vec{R}_{l'})$$

$$= ie^{-ie(\vec{A}_{l'}\cdot\vec{R}_{l'}-\vec{A}_l\cdot\vec{R}_l)/\hbar}\int d^3x \phi^*(\vec{x}-\vec{R}_l)[(\hat{\vec{p}}-e\vec{A})\sum_{l''}U(\vec{x}-\vec{R}_{l''})]\times\hat{\vec{p}}\phi(\vec{x}-\vec{R}_{l'})$$

$$= e^{-ie(\vec{A}_{l'}\cdot\vec{R}_{l'}-\vec{A}_l\cdot\vec{R}_l)/\hbar}\alpha(\vec{L}_{ll'}+\gamma_A\vec{\Gamma}_{ll'}) \qquad (13)$$

where

$$\alpha\vec{L}_{ll'} = i\int d^3x \phi^*(\vec{x}-\vec{R}_l)\hat{\vec{p}}\sum_{l''}U(\vec{x}-\vec{R}_{l''})\times\hat{\vec{p}}\phi(\vec{x}-\vec{R}_{l'}) \qquad (14)$$

$$\alpha\gamma_A\vec{\Gamma}_{ll'} = -i\int d^3x \phi^*(\vec{x}-\vec{R}_l)e\vec{A}\sum_{l''}U(\vec{x}-\vec{R}_{l''})\times\hat{\vec{p}}\phi(\vec{x}-\vec{R}_{l'}) \qquad (15)$$

One will find that $\vec{L}_{ll'}$ is the angular momentum of electrons from orbit motions, while $\gamma_A\vec{\Gamma}_{ll'}$ is the change of angular momentum resulted from the electromagnetic fields. Because there is the factor $\sum_{l''}\vec{\nabla}U(\vec{x}-\vec{R}_{l''})$ in Eq.(14) and the factor is strongly related to the electric polarization, thus $\vec{L}_{ll'}$ is related to the electric polarization and $\vec{L}_{ll'} = \vec{L}_{ll'}(\vec{P})$. $\vec{L}_{ll'}$ should be very small when the polarization vector $\vec{P}=0$, since $\vec{\nabla}U(\vec{x}-\vec{R}_{l''})$ would be directly proportional to the vector $\vec{x}-\vec{R}_{l''}$ for a spherical symmetrical potential. In addition, $\vec{\Gamma}_{ll'} = \vec{\Gamma}_{ll'}(\vec{A})$. Consider the localization of $\phi(\vec{x}-\vec{R}_l)$ and take the "on-site approximation", we obtain

$$H_{SL} = \sum_l 2\alpha\vec{S}_l\cdot(\vec{L}_{ll}+\gamma_B\vec{\Gamma}_{ll}) = \sum_l J_l\vec{S}_l\cdot\vec{L}_l \qquad (16)$$

where $S_l^x = (S_l^+ + S_l^-)/2$, $S_l^y = -i(S_l^+ - S_l^-)/2$, $S_l^z = (S_l^+ - S_l^-)/2$, $S_l^+ = c_{l\uparrow}^+c_{l\downarrow}$, and $S_l^- = c_{l\downarrow}^+c_{l\uparrow}$. These can be written in $\vec{S}_l = c_{l\alpha}^+\vec{\sigma}_{\alpha\beta}c_{l\beta}/2$. $\vec{L}_l$ is the affective angular momentum and $\vec{L}_l = \vec{L}_l(\vec{P},\vec{A})$ as discussed above.

Similarly, the last term in Eq.(1) is

$$H_{ext} = =\sum_l g_l\vec{S}_l\cdot\vec{B}_l \qquad (17)$$

where $\vec{B}_l$ is the magnetic field at $\vec{R}_l$. Synthesize the results above, we arrive at this one band model

$$H = \sum_{l,\delta,\sigma}(te^{-ie\vec{A}_l\cdot\vec{\delta}/\hbar}c_{l\sigma}^+c_{l+\delta\sigma}+\text{h.c.}) + U\sum_l n_{l\sigma}n_{l\bar\sigma} + \sum_{l,l',\sigma,\sigma'}V_{ll'}n_{l\sigma}n_{l'\sigma'} + \sum_l J_l\vec{S}_l\cdot\vec{L}_l + \sum_l g_l\vec{S}_l\cdot\vec{B}_l \qquad (18)$$

where $n_{l\sigma} = c_{l\sigma}^+c_{l\sigma}$ and $\vec{L}_l = \vec{L}_l(\vec{P},\vec{A})$. The term such as $\varepsilon\sum_{l,\sigma}c_{l\sigma}^+c_{l\sigma}$ is neglected since they can be taken in the chemical potential. The model can be extended to other multiband models if it is necessary.

It is obvious that the ferromagnetism (or antiferromagnetism) is included in the model of Eq.(18), because the first and second terms give the well-known Hubbard model if $\vec{A}=0$. The Hubbard model could be deduced to the t-j model, no matter whether this would be yes, the model includes some effects of $\vec{S}_l\cdot\vec{S}_{l'}$ at least. Particularly, the affective $\vec{L}_l$ can play the role of magnetic field, and this is because of the "polarization" of $\vec{L}_l$ when $\vec{\nabla} U(\vec{x} - \vec{R}_{l''})$ has superiority in some direction due to the electric polarization. The ferromagnetism could be steady with the aid of $\vec{L}_l$. Similarly, $\vec{S}_l$ can play the role of electric field, and this is beneficial for a steady macroscopic polarization $\vec{P}$. Therefore, the magnetic and electric orders in multiferroics are strongly coupled and affected each other, and $\sum_l J_l \vec{S}_l \cdot \vec{L}_l$ could be replaced by $\sum_l J_l \vec{S}_l \cdot \vec{p}_l$, where $\vec{p}_l$ is the electric dipole at a site. This explains the field dependent competing magnetic ordering in multiferroic Ni3V2O8[7]. The importance of the spin-orbit interaction has also been noted by many works [8]. However, the polarization of each molecule could not be entirely included in this model although it can show the effect of magnetic order on the electric order; this is because the motion of an electron within a molecule could not be wholly described with any tight binding model. Because the extended Hubbard models include charge orders [9,10], and the forth term in Eq.(18) is beneficial to the ferroelectricity, generally speaking, the ferroelectricity based on correlation effect may be possible due to possible aid of other factors.

Having synthesized the discussion above, we suggest a new model (IJK model)

$$H \sim \sum_{<l,\delta>} I_l \vec{p}_l \cdot \vec{p}_{l+\delta} + \sum_{<l,\delta>} J_l \vec{S}_l \cdot \vec{S}_{l+\delta} + \sum_l K_l \vec{S}_l \cdot \vec{p}_l \qquad (19)$$

The model (19) is not equal to the model (18), while the model (19) may be good for the investigation about phase transition.

In summary, through the derivation of a model, we can arrive at the conclusion that multiferroics could be described with the model shown in Eq.(18) which shows that both ferromagnetism and ferroelectricity are from correlation effects. We believe that many physics will be found in the model.